\documentclass[9pt,twocolumn,twoside]{osajnl_modified}

\journal{optica} 

\setboolean{shortarticle}{false}

\usepackage{lineno}

\title{Soliton linear-wave scattering in a Kerr microresonator}

\author[1,2]{Pierce C. Qureshi}
\author[1,2]{Vincent Ng}
\author[2,3]{Farhan Azeem}
\author[2,3]{Luke S. Trainor}
\author[2,3]{Harald G. L. Schwefel}
\author[1,2]{St\'ephane Coen}
\author[1,2]{Miro Erkintalo}
\author[1,2,*]{Stuart G. Murdoch}

\affil[1]{Department of Physics, University of Auckland, Auckland 1010, New Zealand}
\affil[2]{The Dodd-Walls Centre for Photonic and Quantum Technologies, New Zealand}
\affil[3]{Department of Physics, University of Otago, Dunedin 3016, New Zealand}
\affil[*]{Corresponding author: s.murdoch@auckland.ac.nz}

\begin{abstract}
The nonlinear scattering of a linear optical wave from a conservative soliton has been widely studied in optical fibers as a mechanism for nonlinear frequency conversion. Here we extend this analysis to consider the scattering of an externally injected probe wave from a dissipative Kerr cavity soliton circulating in a Kerr microresonator. We demonstrate, both theoretically and experimentally, that this nonlinear interaction can be harnessed for useful expansion of the soliton frequency comb via the formation of a secondary idler comb. We explore the physics of the process, showing that the phase detuning of the injected probe from a cavity resonance plays a key role in setting the central frequency of the idler comb, thus providing a convenient parameter through which to control the spectral envelope of that comb. Our results elucidate the dynamics that govern the interactions between dissipative Kerr cavity solitons and externally injected probe waves, and could prove useful in the design of future Kerr frequency comb systems by enabling the possibility to provide high-power comb lines in a specified spectral region simply through the injection of a suitably chosen probe.
\end{abstract}

\begin{document}

\maketitle

Microresonator frequency combs (microcombs) offer an attractive new pathway towards the realization of miniaturized, low-power coherent optical frequency combs~\cite{pasquazi_micro-combs_2018, kippenberg_dissipative_2018, gaeta_photonic-chip-based_2019}. Applications of these chip-scale devices have already been demonstrated across fields ranging from telecommunications~\cite{pfeifle_coherent_2014} and spectroscopy~\cite{suh_microresonator_2016} to remote sensing~\cite{trocha_ultrafast_2018,suh_soliton_2018} and medical diagnostics~\cite{marchand_soliton_2021}. In the vast majority of systems, the resonators are driven with a monochromatic laser, and frequency combs form through the third-order optical Kerr nonlinearity~\cite{pasquazi_micro-combs_2018}. In such systems, coherent comb states are underpinned by the excitation of ultra-short localized structures known as temporal cavity solitons (CS)~\cite{leo_temporal_2010} -- also referred to as dissipative Kerr solitons~\cite{herr_temporal_2014, kippenberg_dissipative_2018}.

The spectral properties of Kerr microcombs are dominantly set by the material and geometric parameters of the microresonator. A key comb property is its spectral bandwidth, which (to leading-order) scales as~\cite{coen_universal_2013}
\begin{equation}
\Delta f\propto \sqrt{\frac{\gamma\mathcal{F}}{|\beta_2|}},
\end{equation}
where $\gamma$, $\mathcal{F}$, and $\beta_2$ are the resonator's Kerr nonlinearity coefficient, finesse, and group-velocity dispersion (GVD) at the driving wavelength, respectively. The large $\mathcal{F}$ and $\gamma$ of microresonators are thus highly conducive to the realization of combs with large spectral bandwidth. In addition, the dispersion of microresonators can be engineered to provide further increase in the attainable comb bandwidth. This typically involves both lowering the resonator's GVD coefficient $|\beta_2|$, and optimizing its third-order dispersion so as to allow additional spectral extension through the formation of dispersive waves (DW) that are phase-matched to the pump~\cite{jang_observation_2014,brasch_photonic_2016,milian_soliton_2014}.

Very recently, a new approach has been put forward that allows for the spectral  characteristics and extent of microcombs to be controlled even after the fabrication of the resonator (whereupon its material and geometric properties are set). Specifically, the injection of a frequency-shifted probe field alongside the main comb generating pump has been shown to permit controllable expansion of the comb spectrum via two distinct mechanisms~\cite{zhang_spectral_2020, zhang_dark-bright_2021,moille_ultra-broadband_2021}. In the first mechanism~\cite{zhang_spectral_2020, zhang_dark-bright_2021}, a CS excited by the pump field imparts a nonlinear phase shift on the intracavity field of the frequency-shifted probe. This induces a frequency comb structure around the probe frequency via cross-phase modulation (XPM). In the second mechanism, frequency components of the soliton comb are spectrally translated through the four-wave-mixing process of nonlinear Bragg scattering~\cite{moille_ultra-broadband_2021}, again giving rise to a comb structure around the frequency-shifted probe field. Both mechanisms result in the generation of a secondary comb around the probe frequency, with line spacing equal to the spacing of the original soliton comb, yet they rely on fundamentally different physical phenomena: incoherent XPM~\cite{zhang_spectral_2020, zhang_dark-bright_2021} and coherent Bragg scattering~\cite{moille_ultra-broadband_2021}. Here we focus on the latter mechanism.

Recent experiments have unequivocally demonstrated the application potential of Bragg-scattering spectral expansion: a coherent Kerr microcomb with an expanded bandwidth of 1.6 octaves was reported in~\cite{moille_ultra-broadband_2021}. However, a number of questions remain open with regards to the physics that underpin the phenomenon. For instance, earlier studies~\cite{xu_cascaded_2013, webb_nonlinear_2014} have shown that nonlinear Bragg scattering of a frequency comb can be understood as the frequency-domain description of a particular soliton-linear wave (LW) interaction~\cite{yulin_four-wave_2004, efimov_time-spectrally-resolved_2004, skryabin_theory_2005, efimov_interaction_2005} that has been extensively studied in the context of conservative (single-pass) nonlinear fibre optics~\cite{philbin_fiber-optical_2008, demircan_controlling_2011, bendahmane_observation_2015, ciret_observation_2016}. This raises the question as to whether the simple phase-matching conditions known to govern the process in the single-pass case hold predictive power in the resonator context. In addition, nonlinear Bragg scattering is intrinsically a coherent FWM process, and could therefore be envisaged to depend upon the linear detuning of the probe wave from a cavity resonance; however, no discussion of the role of detuning has hitherto been presented.

Here we theoretically and experimentally study the spectral extension of soliton microcombs via coherent FWM Bragg-scattering. We show that the simple phase-matching condition that underpins soliton-LW interactions in single-pass systems~\cite{webb_nonlinear_2014} remains valid in the resonator context, provided however that the phase detuning of the probe wave is appropriately accounted for. Indeed, we find that this detuning plays a key role in the process, providing a convenient parameter through which to control the spectral characteristics of the comb extension. We perform experiments in a magnesium-fluoride ($\text{MgF}_2$) micro-disk resonator, and demonstrate the generation of frequency tunable, low-noise idler combs that possess identical line spacing, and the same low-noise characteristics, as the driving CS comb.  We further highlight the flexibility of this comb expansion technique by swapping the spectral locations of the CS pump and the probe wave whilst still maintaining the required phasematching -- and hence comb extension. Our results provide significant insights into the spectral extension of soliton microcombs, and highlight the intimate linkage between Bragg scattering spectral extension and soliton-LW interactions that have been studied widely in the context of nonlinear fiber optics and supercontinuum generation~\cite{xu_cascaded_2013, webb_nonlinear_2014,yulin_four-wave_2004, efimov_time-spectrally-resolved_2004, skryabin_theory_2005, efimov_interaction_2005, philbin_fiber-optical_2008, demircan_controlling_2011, bendahmane_observation_2015, ciret_observation_2016}.

We begin by briefly recounting how interactions between solitons and weak linear waves can enact resonant energy transfer to new frequencies in Kerr media~\cite{yulin_four-wave_2004, efimov_time-spectrally-resolved_2004, skryabin_theory_2005, efimov_interaction_2005,wai_nonlinear_1986, akhmediev_cherenkov_1995, skryabin_colloquium_2010, erkintalo_pump-soliton_2011}. Considering a superposition field $E = S+p$, where $S$ and $p$ represent the soliton and the probe, respectively, the Kerr nonlinearity $|E|^2E$ will after linearization with respect to $p$ yield three terms: $|S|^2S$, $2|S|^2p$, and $S^2p^\ast$. Each of these three terms can drive resonant energy transfer to new waves~\cite{skryabin_colloquium_2010}, provided that those waves are phase-matched to one of the driving terms. In fact, the first term is responsible for the generation of the standard DWs that are directly phasematched with the soliton~\cite{wai_nonlinear_1986, akhmediev_cherenkov_1995}. On the other hand, the second and third terms represent nonlinear mixing between the soliton and the weak probe wave, and can drive additional phasematched radiation processes~\cite{yulin_four-wave_2004, efimov_time-spectrally-resolved_2004, skryabin_theory_2005, efimov_interaction_2005,erkintalo_pump-soliton_2011}. In particular, the $2|S|^2p$ interaction has been shown to drive both the incoherent XPM and coherent FWM Bragg scattering interactions~\cite{skryabin_theory_2005}.

\begin{figure}[!t]
\centering
  \includegraphics[width = \linewidth, clip = true]{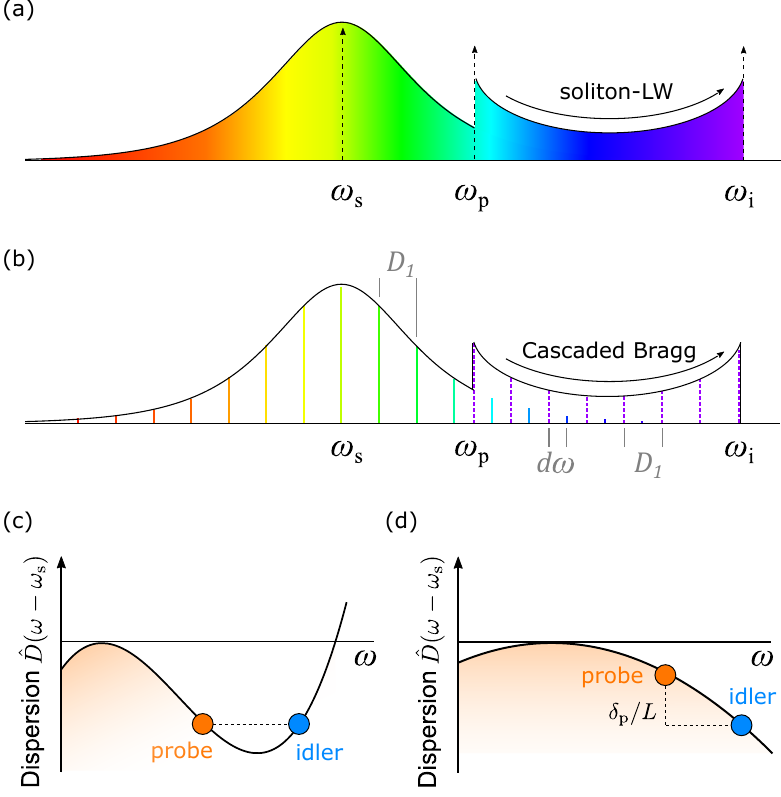}
  \caption{(a) Schematic illustration of the interaction between a soliton at $\omega_\mathrm{s}$ and a weak linear probe wave at $\omega_\mathrm{p}$. The interaction enables the flow of energy to a phase-matched idler wave at $\omega_\mathrm{i}$. (b) Frequency domain description of the soliton-LW interaction in (a), showing how the energy flow can be described as a cascade of nonlinear Bragg-scattering events. Components of the newly-generated idler comb are all shown in purple for clarity. The idler comb possesses the same $\text{FSR}=D_1/(2\pi)$ but is offset from the soliton comb (by $d\omega$). (c) and (d) visualize the phasematching of the process in single-pass and resonator configurations, respectively. (c) In single-pass configurations, higher-order dispersion is required for phasematching. (d) The extra degree of freedom provided by the probe detuning $\delta_\mathrm{p}$ relaxes the phasematching condition in resonators.}
  \label{fig1}
\end{figure}

Earlier studies have shown that the soliton-LW interaction can be understood in the frequency domain as a cascade of individual FWM Bragg scattering processes~\cite{xu_cascaded_2013, webb_nonlinear_2014}. Specifically, pairs of discrete frequency components of a periodic train of solitons can act as pumps that drive a Bragg scattering cascade that translates the incident LW to a new idler frequency [see Fig.~\ref{fig1}(a) and (b)]. Remarkably, this cascade can be phase-matched even though none of the elementary FWM processes are phase-matched~\cite{xu_cascaded_2013, webb_nonlinear_2014, erkintalo_cascaded_2012}; moreover, the phase-matching condition of the entire cascade is (approximately) the same as that of the soliton-LW interaction driven by the nonlinear polarization term $2|S|^2p$. In what follows, we show that this result extends to resonator configurations by demonstrating that the spectral extension of soliton microcombs via FWM Bragg scattering obeys the phasematching condition of the pertinent time-domain soliton-LW interaction.

\begin{figure*}[!t]
\centering
  \includegraphics[width = \textwidth, clip = true]{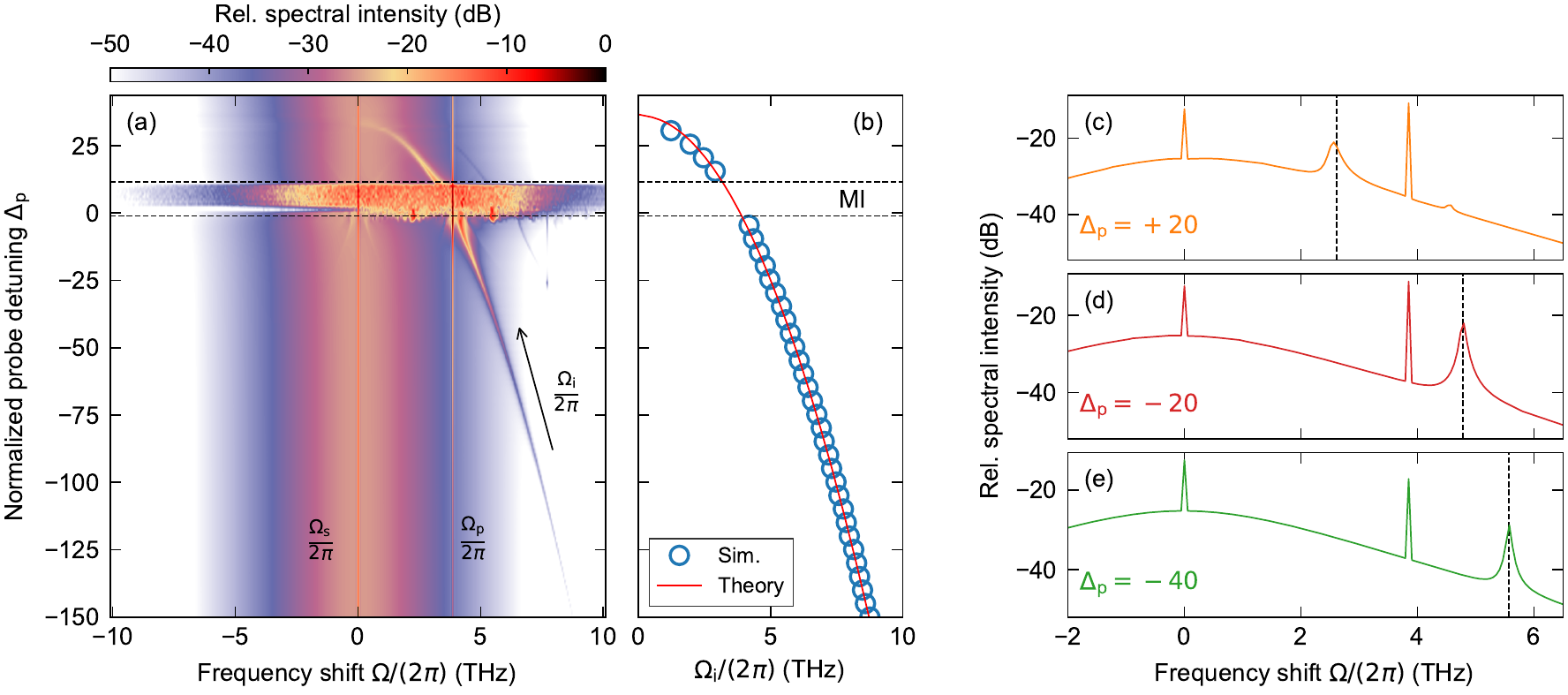}
  \caption{Numerical simulation results that illustrate soliton-LW interactions in a Kerr microresonator. (a) Pseudo-color plot shows the intracavity spectrum as a function of the probe detuning normalized to half the cavity linewidth, $\Delta_\mathrm{p}=\delta_\mathrm{p}/\alpha$. The $x$-axis denotes the frequency offset from the soliton-generating pump, $f=(\omega-\omega_\mathrm{s})/(2\pi)$. $\Omega_\mathrm{s}$, $\Omega_\mathrm{p}$, and $\Omega_\mathrm{i}$ respectively denote angular frequency detunings of the soliton ($\Omega_\mathrm{s} = 0$), the probe, and the idler. (b) Circles show the peak frequency detuning of the idler comb as a function of the linear probe detuning $\Delta_\mathrm{p}$ as extracted from the simulation data in (a). Solid red curve in (b) shows the phasematched frequency predicted by Eq.~\eqref{pm2}. Dashed horizontal lines in (a) and (b) delineate the region within which the probe undergoes modulational instability (MI) and prohibits the soliton-LW interaction. (c)--(e) show individual spectra from (a) at selected probe detunings as indicated, with dashed vertical lines indicating the phasematched idler frequencies predicted by Eq.~\eqref{pm2}. Simulation parameters are similar to the experiments that will follow: $\mathcal{F}=3.1\times10^4$, $\gamma = 1.4\times10^{-3}~\mathrm{W^{-1}m^{-1}}$, $\beta_2 = -3.2\times10^{-27}~\mathrm{s^{2}m^{-1}}$, $P_\mathrm{in,i}=|E_\mathrm{in,i}|^2=80~\mathrm{mW}$, with $\mathrm{i}=(\mathrm{s,p})$, and $\theta = 1\times10^{-4}$.}
  \label{fig2}
\end{figure*}

To derive the soliton-LW phasematching condition in a resonator configuration, we note that the phase accumulated by the linear idler wave at frequency $\omega_\mathrm{i}$ over one round trip is: $\phi_\mathrm{i} = \beta(\omega_\mathrm{i})L-\omega_\mathrm{i}t_\mathrm{R}$, where $\beta(\omega)$, $L$, and $t_\mathrm{R}$ are respectively the propagation constant, length, and round trip time of the resonator. Phase-matching is achieved when $\phi_\mathrm{i} = \phi_\mathrm{p}+ 2\pi m$, where $\phi_\mathrm{p}=-\omega_\mathrm{p}t_\mathrm{R}$ is the phase accumulated by the externally-injected probe wave (at the fixed coupling position) and $m$ is an integer~\cite{luo_resonant_2015}. Expanding the propagation constant $\beta(\omega)$ as a Taylor-series around the soliton pump, $\omega_\mathrm{s}$, we obtain
\begin{equation}
\hat{D}(\omega_\mathrm{i}-\omega_\mathrm{s})L = \delta_\mathrm{p}+\hat{D}(\omega_\mathrm{p}-\omega_\mathrm{s})L + 2\pi q,
\label{pm}
\end{equation}
where $\delta_\mathrm{p}$ is the phase detuning of the injected probe from the cavity resonance closest to it, and the reduced dispersion
\begin{align}
\hat{D}(\omega-\omega_\mathrm{s}) &= \beta(\omega)-\beta(\omega_\mathrm{s})-\frac{t_\mathrm{R}}{L}(\omega-\omega_\mathrm{s}) \\
& = \sum_{k\geq 2} \frac{\beta_k}{k!}(\omega-\omega_\mathrm{s})^k,
\end{align}
The coefficient $q=m-m_0$ with $m_0$ the mode index of the resonance closest to $\omega_\mathrm{p}$ describes phase-matching to higher-order resonant sidebands that will not be considered in this work; in what follows, we set $q = 0$.

Equation~\eqref{pm} is akin to the phasematching condition for Bragg-type soliton-LW interactions in single-pass fibre configurations~\cite{webb_nonlinear_2014}, but with the additional probe detuning $\delta_\mathrm{p}$ accounting for the resonant nature of the system. Importantly, the presence of $\delta_\mathrm{p}$ significantly relaxes the phasematching requirements: whilst higher-order dispersion is required in single-pass configurations, in resonators phasematching can be achieved even with a purely quadratic dispersion profile [see Fig.~\ref{fig1}(c) and (d)]. Explicitly, for a resonator with second order dispersion only, the (angular) frequency detuning $\Omega_\mathrm{i} = \omega_\mathrm{i}-\omega_\mathrm{s}$ of the newly generated idler field from the soliton pump satisfies:
\begin{equation}
\Omega_\mathrm{i}^2 = \Omega_\mathrm{p}^2 + \frac{2\delta_\mathrm{p}}{\beta_2 L},
\label{pm2}
\end{equation}
where $\Omega_\mathrm{p} = \omega_\mathrm{p}-\omega_\mathrm{s}$. It is worth noting that Eq.~\eqref{pm2} predicts two phasematched idler frequencies symmetrically located either side of the CS pump. However, the FWM Bragg cascade to the idler frequency located on the opposite side of the CS to the probe will require substantially more steps and hence be less efficient. Indeed, for the parameters used in this paper, only the idler frequency closest to the probe frequency is observed in simulations and experiments. For different resonator parameters, however, it may be possible to efficiently drive both cascades resulting in new spectral components generated on both sides of the CS~\cite{moille_ultra-broadband_2021}.

To confirm the analysis above, we first perform numerical simulations using the generalized Lugiato-Lefever equation (LLE)~\cite{zhang_spectral_2020, coen_modeling_2013, chembo_spatiotemporal_2013, hansson_bichromatically_2014, taheri_optical_2017}:
	\begin{equation}
		\begin{split}
		t_R\frac{\partial E(t,\tau)}{\partial t}=\left[ -\alpha-i\delta_\mathrm{s}-i\frac{\beta_2 L}{2} \frac{\partial^2}{\partial \tau^2}+i\gamma L|E|^2\right]E+\sqrt{\theta}E_\mathrm{in}(t,\tau).
		\end{split}
		\label{LLE}
	\end{equation}
Here $E(t,\tau)$ describes the slowly-varying intracavity electric field envelope, $t$ is a ``slow'' time that describes the evolution of $E(t,\tau)$ at the scale of the photon lifetime, $\tau$ is a corresponding ``fast'' time that describes the envelope's profile over a single roundtrip, $\alpha=\pi/\mathcal{F}$ is half of the power lost per round trip (and equal to the resonance half-width), $\delta_\mathrm{s}$ is the phase detuning of the soliton-generating pump from the cavity resonance closest to it, and $\theta$ is the input power coupling coefficient. The driving field $E_\mathrm{in}(t,\tau)$ accounts for the bi-chromatic pumping and is defined as
\begin{equation}
E_\mathrm{in}(t,\tau) = E_\mathrm{in,s} + E_\mathrm{in,p}e^{-i\Omega_\mathrm{p}\tau + i\left(\delta_\mathrm{p}-\delta_\mathrm{s}+\frac{\beta_2L}{2}\Omega_\mathrm{p}\right) \frac{t}{t_R}},
\end{equation}
where $E_\mathrm{in,s}$ and $E_\mathrm{in,p}$ are the pump amplitudes with units of $\mathrm{W}^{1/2}$ at $\omega_\mathrm{s}$ and $\omega_\mathrm{p}$, respectively.

\begin{figure*}[!t]
\centering
  \includegraphics[width = \textwidth, clip = true]{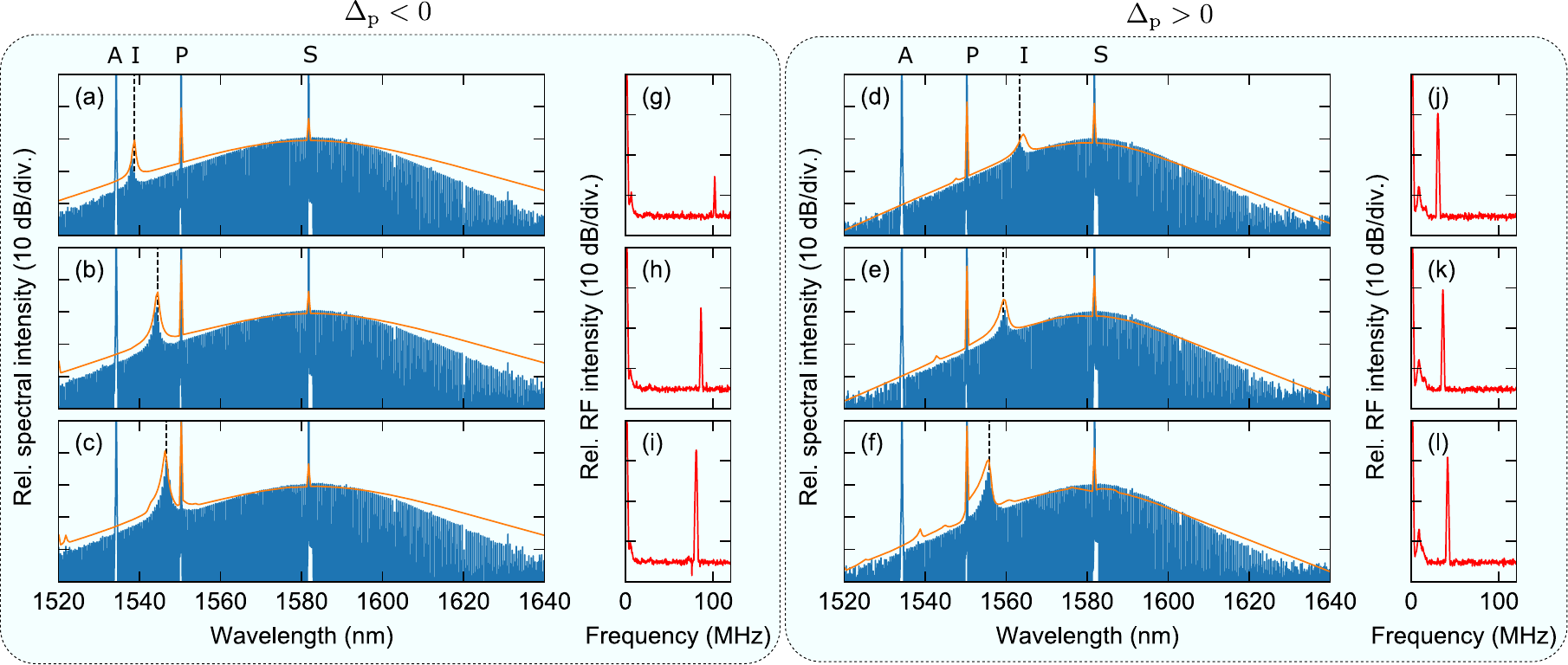}
  \caption{Experimental results with the soliton-generating pump at $1582~\mathrm{nm}$ and the probe at $1550~\mathrm{nm}$. (a)--(c) Blue curves show microcomb spectra measured at the resonator output for different \emph{negative} values of the probe detuning, such that the idler is generated at wavelengths \emph{shorter} than the probe.  Orange curves show corresponding results from simulations and the dashed vertical lines show the phasematched wavelength predicted by Eq.~\eqref{pm2}. (d)--(f) show spectra as in (a)--(c) but for different \emph{positive} values of the probe detuning, such that the idler is generated at wavelengths \emph{longer} than the probe. (g)--(l) Low-frequency RF spectra corresponding to (a)--(f). The beat frequencies in the RF spectra correspond to the offset between the soliton and the idler combs, $\delta\omega/(2\pi)$. Labels A, I, P, and S on the top respectively indicate the auxiliary laser used for thermal compensation, the idler, the probe, and the soliton.}
  \label{fig3}
\end{figure*}

Figure~\ref{fig2} shows results from numerical simulations that use parameters similar to the experiments that will follow (see figure caption). In particular, the false color plot in Fig.~\ref{fig2}(a) shows the spectral envelope of the intracavity field as a function of the probe phase detuning $\delta_\mathrm{p}$ normalized to the resonance half-width ($\Delta_\mathrm{p} = \delta_\mathrm{p}/\alpha$), whilst Figs.~\ref{fig2}(c)--(e) show lineplots of selected spectra in more detail. At each probe detuning, a CS is first excited (at $\Omega = \omega-\omega_\mathrm{s} = 0$) and allowed to stabilize before the probe field ($E_\mathrm{in,p}$) is turned on.

In addition to the broadband soliton field around $\Omega = 0$ and the injected probe field at $\Omega_\mathrm{p}$, a third wave whose position changes continuously with the probe detuning is clearly visible. This corresponds to the idler comb that is generated via the soliton-LW interaction described above. Indeed, in Fig.~\ref{fig2}(b) we plot the peak frequency shift of the idler comb as a function of the probe detuning, superimposed with the theoretical prediction of the phasematched frequency given by Eq.~\eqref{pm2}. The agreement between the simulated idler frequency and the theoretical prediction is almost perfect. When the probe detuning is sufficiently close to its own resonance [with $\Delta_\mathrm{p}\in (-3,11)$], our simulations predict that the soliton-LW interaction is interrupted. This occurs because the probe itself undergoes modulational instability (MI) within this range of detunings, as clearly evidenced by the characteristic MI spectral features observed in Fig.~\ref{fig2}(a). For most detunings within this range, the large intensity fluctuations of MI prevent the persistence of the CS altogether~\cite{anderson_coexistence_2017, nielsen_coexistence_2019} and accordingly prohibit any soliton-LW interaction. However, for a narrow sliver of detunings [with $\Delta_\mathrm{p}\in (-3,1)$], our simulations predict that the CS can coexist with the MI state associated with the probe, and in this case a LW interaction can take place, but with the noise from the probe's MI transferred to both the soliton and the idler combs, resulting in low-coherent states. This is in contrast with the rest of the detunings shown in Fig.~\ref{fig2}, where the simulations confirm that both the soliton and idler combs are coherent as required for useful spectral extension.

For experimental demonstration, we use a setup that is built around an $\text{MgF}_2$ micro-disk shaped via diamond point turning and then hand-polished to achieve a measured finesse of $\mathcal{F} \approx 3.1\times10^4$ at 1550~nm~\cite{grudinin_ultra_2006,strekalov_nonlinear_2016,sayson_octave-spanning_2019}. The disk has a major radius of $600~\mu\mathrm{m}$, yielding an FSR of 58.4~GHz. A micron diameter fiber taper is used to couple the driving fields to and from the resonator. The micro-disk is driven by three optical fields derived from two C-band external-cavity-lasers (ECL), and one L-band ECL. Each laser is amplified by an erbium-doped-fiber amplifier, then filtered and coupled to the optical taper using fiber wavelength-division-multiplexers. One of the C-band lasers and the L-band laser are coupled to the same mode family, and they can interchangeably act as the soliton-generating pump and the probe field. The remaining C-band laser (wavelength fixed at 1534~nm) is coupled to a different mode-family of the resonator and configured to act as an auxiliary pump that provides thermal compensation during the CS excitation~\cite{zhang_sub-milliwatt-level_2019, lu_deterministic_2019} -- this laser does not participate in the soliton-LW dynamics. Throughout our measurements, the powers of the soliton and probe lasers were set to 80~mW whilst that of the auxiliary pump was set to 150~mW.

We first set the wavelength of the L-band pump to 1582~nm and excite a CS at that wavelength. The remaining C-band laser acts the probe, and is tuned into resonance close to 1550~nm from the blue-detuned (negative $\Delta_\mathrm{p}$) side. Figures~\ref{fig3}(a)--(c) show the comb spectra measured at the output of the fiber taper as the probe detuning is scanned into resonance. We can clearly observe an idler peak that moves towards the probe as the detuning $\Delta_\mathrm{p}$ is increased. The idler peak tunes continuously until we reach the MI region of the probe field, at which point the CS ceases to exist and the soliton-LW interaction is halted -- in accordance with our simulations [see Fig.~\ref{fig2}(a) and (b)]. We then reset the experiment such that a new CS is excited at 1582~nm, and tune the probe field into resonance from the red-detuned (positive $\Delta_\mathrm{p}$) side. Figures~\ref{fig3}(d)--(f) show the resulting spectra, and we again observe the appearance of an idler peak that continuously tunes towards the probe, now as $\Delta_\mathrm{p}$ is decreased. These results are in qualitative agreement with the phasematching prediction of Eq.~\eqref{pm2}, with positive (and negative) probe detunings observed to generate idler peaks at frequencies that are below (and above) the original probe frequency.

Figures~\ref{fig3}(g)--(l) show the low-frequency RF spectra of each microcomb state shown in Fig.~\ref{fig3}(a)--(f). All of the spectra are clean with no excess noise, indicating that we are indeed observing low-noise, stable frequency combs. Yet, each RF spectrum does exhibit a single sharp RF tone. This is because, whilst the idler comb inherits the line spacing of the soliton comb, the two are offset from one another [see Fig.~\ref{fig1}(b)], resulting in beating between adjacent lines in the region where the combs overlap. The beat frequency corresponds to the separation between the probe frequency and the soliton comb line closest to it, and can be written in terms of the soliton and probe detunings and the dispersive shift of the probe resonance as:
\begin{equation}
d\nu \approx \left[\delta_\mathrm{p} - \delta_\mathrm{s} + \frac{\beta_2L\Omega_\mathrm{p}^2}{2}\right]\frac{\text{FSR}}{2\pi}.
\end{equation}
In our experiments, the RF beat tones range from 30 -- 140~MHz, and are hence far too small to be resolved optically. Indeed, the optical comb spectra shown in Fig.~\ref{fig3} were recorded at an optical spectrum analyzer (OSA) resolution of 5 GHz, and a careful examination of these traces reveals only a single set of equi-spaced comb lines. Likewise, further RF intensity-noise measurements made using an electronic spectrum analyzer reveal only a low-noise background out to a frequency of 10 GHz (in addition to the low-frequency RF beat-notes already observed in Fig.~\ref{fig3}). Combined, these measurements confirm that the output spectrum is indeed composed of two individual low-noise combs: the first, the CS comb, and the second, the idler comb extending from the probe frequency to the phasematched idler peak, and spectrally offset from the CS comb.

The ability to measure the RF beat frequency between the soliton and the idler comb allows the two combs to be ``stitched'' together, i.e., the absolute frequencies of the idler comb lines to be directly related to the frequencies of the CS comb. In addition, when combined with measurements of the spectral position of the idler wave, the beat frequencies allow us to accurately estimate the parameters of the entire experiment. First, the change in probe detuning between each spectrum shown in Fig.~\ref{fig2} can be calculated from the change in the measured RF beat notes, $\Delta\delta\nu$, as $\Delta\delta_\mathrm{p}=2\pi\cdot(\Delta \delta\nu/\text{FSR})$; second, the GVD coefficient $\beta_2$ can be extracted by fitting the change in idler position with the change in probe detuning: $\Delta\Omega_\mathrm{i}=(\beta_2L\Omega_\mathrm{i})^{-1}\Delta\delta_\mathrm{p}$; third, $\gamma$ and $\delta_\mathrm{s}$ can be estimated by fitting numerical simulations to the CS spectrum, and to the measured value of $\delta_\mathrm{p}$ at which the probe undergoes MI. These procedures yield the parameters quoted in the caption of Fig.~\ref{fig2}, and when used in LLE simulations, provide results that are in excellent agreement with our experiments, as shown by the orange solid curves in Fig.~\ref{fig3}(a)--(f). We observe particularly how the position and magnitude of the newly generated idler combs are very well reproduced by the simulations, and we also note that the idler positions are well predicted by the phasematched frequency given by Eq.~\eqref{pm2}.

\begin{figure}[!t]
\centering
  \includegraphics[width = \columnwidth, clip = true]{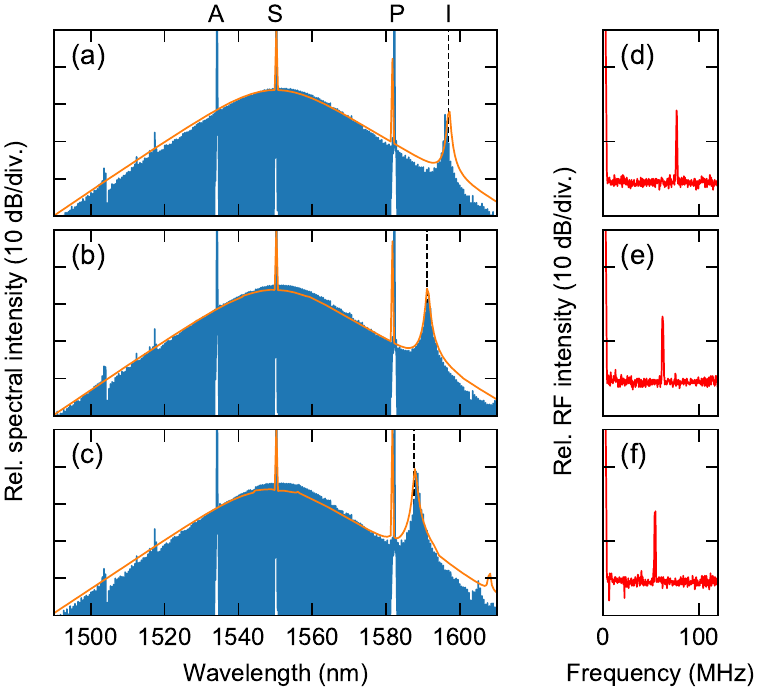}
  \caption{Experimental results as in Fig.~\ref{fig3} but with the soliton and probe waves interchanged: the soliton is here generated at 1550~nm whilst the probe sits at 1582~nm. The results shown are for different negative values of the probe detuning.}
  \label{fig4}
\end{figure}

The soliton and probe fields are interchangeable: we can exchange their roles by selecting appropriate detunings and still generate a phase-matched idler comb. To show this, we set the detuning of the 1550 nm laser such that it generates a CS, and use the 1582 nm laser as the external probe. Figures~\ref{fig4}(a)--(c) show the output comb spectra as the probe field is continuously tuned into resonance from the blue-detuned (i.e., negative $\Delta_\mathrm{p}$) side, whilst Figs.~\ref{fig4}(d)--(f) show the corresponding RF spectra. The observed spectra are qualitatively identical to those obtained when the 1582~nm pump generates the CS. Moreover, we again find excellent agreement between the experimentally recorded spectra and the numerical results obtained from the LLE (red traces). The only adjustment to the simulation parameters used in Figs.~\ref{fig2} and~\ref{fig3} was a change to the value of $\beta_2=-2.9\times10^{-27}~\mathrm{s^{2}m^{-1}}$ due to the change in pump wavelength and residual third-order dispersion.

\begin{figure}[!t]
\centering
  \includegraphics[width = \columnwidth, clip = true]{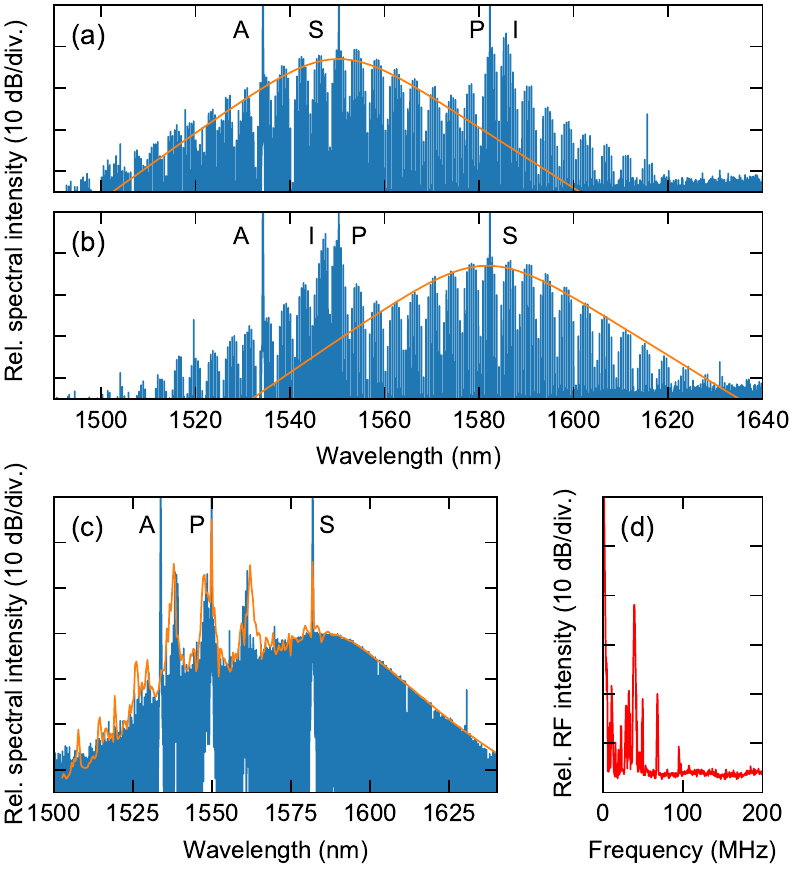}
  \caption{Blue curves in (a) and (b) show experimentally measured spectra at the resonator output when a two-CS state interacts with a probe. In (a) the soliton is at 1550~nm and the probe at 1583~nm, whilst in (b) the roles are swapped. The orange curves in (a) and (b) show theoretically predicted CS envelopes in the absence of the probe wave, highlighting how the interaction leads to substantial spectral extension. (c) Blue and orange curves respectively show experimental and simulated optical spectra in the narrow parameter regime where the probe at 1550~nm undergoes MI yet permits the CS at 1582~nm to persist. (d) shows the low-frequency RF spectrum corresponding to (c), highlighting the low-coherence of the state.}
  \label{fig5}
\end{figure}

The results presented in Figs.~\ref{fig3} and~\ref{fig4} offer strong experimental support for the soliton-LW phasematching theory presented in the previous section. In addition to dynamics involving a single CS and the probe wave, our experiments show that more complex interactions can also take place. First, the soliton-LW interaction can occur when the resonator hosts several solitons. Indeed, in Fig.~\ref{fig5}(a) we show comb spectra recorded at the resonator output with two temporally separated CSs centred around a 1550~nm pump interacting with a probe field at 1582~nm; Fig.~\ref{fig5}(b) shows the corresponding spectrum with the roles of the CS-generating pump and the probe reversed. In both cases, a strong spectral modulation with a period of 4~nm can be observed, indicating the presence of two CS in the cavity separated by 2~ps. This spectral modulation manifests itself across both the CS and the idler combs, thus providing further evidence that the two combs are temporally locked to each other. Also shown in Figs.~\ref{fig5}(a) and (b) are theoretical $\text{sech}^2$ profiles of the CS spectral envelope in the absence of the probe wave. This shows how the soliton-LW interaction leads to a substantial increase in comb intensity around the probe. Finally, in accordance with the simulations shown in Fig.~\ref{fig2}, our experiments show that a narrow region of probe detunings exist where the probe can undergo modulational instability whilst still allowing the CS and the idler comb to persist. Figure~\ref{fig5}(c) shows the measured optical spectrum recorded in this regime, superimposed with corresponding simulations, and we indeed observe characteristic (i) MI sidebands around the 1550~nm probe wavelength and (ii) soliton $\text{sech}^2$ profile around the 1582~nm CS pump. Moreover, the RF spectrum of this state [see Fig.~\ref{fig5}(d)] shows clearly elevated level of intensity noise, as expected due to the chaotic nature of the MI state.

To summarize, we have shown that the nonlinear interaction between a cavity soliton and an externally injected probe can engender spectral extension of a Kerr microcomb via the formation of a secondary idler comb. We have shown that the process is underpinned by a simple phasematching condition derived from the time-domain soliton-LW interaction picture, revealing that the probe field's linear detuning plays a key role in controlling the phasematched frequency -- and hence the spectral characteristics of the idler comb. The idler and its generating soliton comb have the same line spacing and low-noise characteristics, but are spectrally offset from one another; measurement of the RF beat frequency between the two combs allows the frequencies of the idler comb lines to be directly related to the CS comb. Combined, these properties make soliton-LW scattering an attractive candidate for the spectral expansion of Kerr combs, enabling comb power to be enhanced at desired spectral locations simply through the injection of an appropriate external probe field. We anticipate this ability could find useful application in many areas of microcomb research, including spectroscopy, frequency metrology and optical frequency synthesis. Finally, we close by emphasizing that the soliton-LW interaction described in our work represents the time-domain description of cascaded FWM Bragg scattering~\cite{xu_cascaded_2013, webb_nonlinear_2014}, which has been linked to microcomb spectral extension in recent works completed in parallel with our study~\cite{moille_ultra-broadband_2021}.

\section*{Funding}
Marsden Fund and the Rutherford Discovery Fellowships of the Royal Society of New Zealand.

\section*{Disclosures}
\noindent The authors declare no conflicts of interest.

\section*{Data availability}
\noindent The data that support the plots within this paper and other findings of this study are available from the corresponding author upon reasonable request.

\bibliography{Pierce}

\end{document}